\documentclass[12pt, draftclsnofoot, onecolumn]{IEEEtran}
\usepackage{blindtext}

\ifCLASSINFOpdf
\else
\fi

\usepackage{amssymb}
\usepackage{amsmath}
\usepackage[T1]{fontenc}
\usepackage{graphicx}
\usepackage{caption}
\usepackage{subcaption}
\usepackage{cite}
\usepackage{float}
\usepackage{authblk}
\usepackage{mathtools}
\usepackage{siunitx}

\title{Diffusion-based Molecular Communication Channel in Presence of a Probabilistic Absorber: Single Receptor Model and Congestion Analysis}

\author[1]{Shirin Salehi}
\author[1]{Naghmeh S. Moayedian}
\author[2]{Eduard Alarcon}
\affil[1]{\small Department of Electrical and
Computer Engineering, Isfahan University of Technology, Isfahan 84156-
83111, Iran}
\affil[2]{\small NaNoNetworking Center in Catalunya, Universitat
Polit\`ecnica de Catalunya, 08034 Barcelona, Spain
}

\begin{document}

\maketitle

\begin{abstract}
In this paper, a diffusion-based molecular communication channel is modeled in presence of a probabilistic absorber. The probabilistic absorber is an absorber which absorbs molecules upon collision with probability $q$. With random walk analysis, the discrete probability function of particle location in presence of a probabilistic absorber can be found. Then a continuous probability function is fitted to this Markov-based results with introducing several fitting parameters to the known probability function of particle location in an unbounded environment without an absorbing barrier. With this approach, a single receptor is modeled as an M/M/1/1 queue in which $q$ represents the complementary blocking probability and the mean service time is the mean trafficking time.
Therefore, we are able to model the stochastic nature of ligand-receptor binding which comes from the incapability of a receptor to receive all molecules in its space; Also known as receptor occupancy. 
Proper consideration of the absorption effect leads to the accurate calculation of the concentration at the desired site, which is definitely less than the concentration obtained when neglecting it. These findings can have a crucial role in designing drug delivery systems in which determining the optimal rate of the drug transmitting nanomachines is critical to avoid toxicity while maintaining effectiveness.
\end{abstract}

\begin{IEEEkeywords}
Molecular communication, channel impulse response, probabilistic absorber, diffusion, random walk, Markov chain, receptor modeling.
\end{IEEEkeywords}

\section{Introduction}
Molecular communication (MC) is a new communication paradigm in which the exchange of information happens through messenger molecules in a fluid medium. MC is known as the most practical way of communication between nanomachines due to the compliance with scale and environment~\cite{akyildiz2008nanonetworks}. The advantages of this communication scheme over nanoelectromagnetic communication are its intrinsic nanoscale characteristic, potential biocompatibility, and low energy consumption~\cite{gine2009molecular}. In MC, chemical signals or molecules are transmitted and received. This way of communication can be absolutely suitable for medical purposes~\cite{felicetti2016applications,chahibi2013molecular,chude2017molecular,nakano2013transmission,salehi2017releasing,chahibi2017molecular}.
\begin{figure*}
\centering
\includegraphics[width=0.9\textwidth]{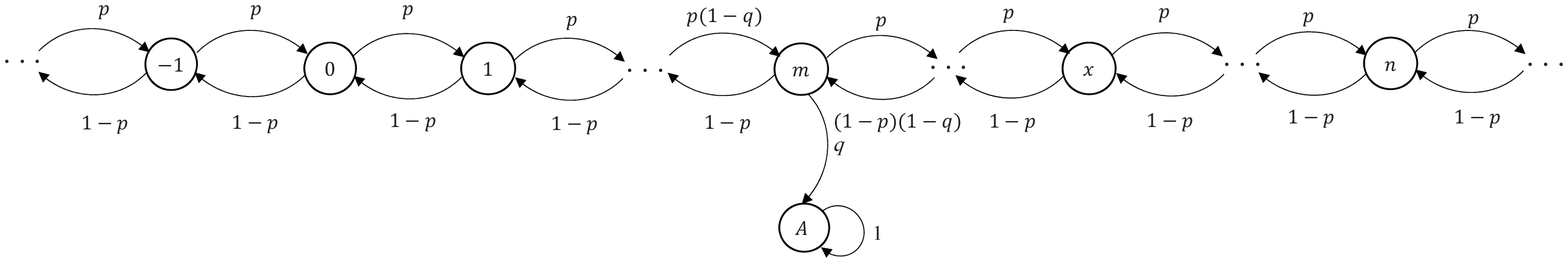}
\caption{Markov chain representation of 1-D random walk with a probabilistic absorber located at $m$, observation point at $x$}
\label{markovchain}
\centering
\end{figure*}
MC components include Transmitting nanomachine (TN), Receiving nanomachine (RN), messenger molecules, interface molecules and guiding and transport mechanisms. There are five known mechanisms of guiding and transport in MC~\cite{atakan2016molecular} including diffusion with or without drift, gap junction channels, molecular motors, self-propelling microorganisms and random collision of mobile nanomachines. 

Among the mentioned guiding and transport mechanisms, free diffusion is of great importance. Free diffusion is suitable for dynamic and unpredictable environments where no infrastructure is available for MC. It also takes advantage of zero energy consumption during propagation. However, this transport mechanism requires a large number of molecules. Furthermore, due to the random movement of particles, the required time to get to the destination can be significant~\cite{nakano2013molecular,atakan2016molecular}.

Determining the molecular channel impulse response is necessary to design an MC system including equalization and detection. The MC channel impulse response is defined as the expected number of molecules counted at a particular location due to an instantaneous emission by a TN located at the origin of the coordinate system at time $t=0$. This is often denoted by the concentration function, $C(X,t)$, in which $X$ is coordinate of the location under consideration and $t$ is the time duration after the instantaneous release. 

A widely employed approach to determine the channel impulse response in MC is solving Fick's second law of diffusion and applying the initial and boundary conditions of the problem~\cite{atakan2016molecular,crank1979mathematics}. However, this approach has several shortcomings e.g. solving Fick's second law of diffusion is only possible with considering simplifying assumption for the problem under consideration such as assuming a point source and an unbounded environment~\cite{jamali2016channel}. 
The topic of channel impulse response has been frequently addressed in MC literature. Each contribution attempts to calculate the diffusion channel impulse response with lifting one of the common simplifying assumptions in the field of MC and derive a more accurate channel model. In~\cite{pierobon2010physical}, one of the first conducted researches in the field of diffusion-based molecular communication (DMC), a physical end-to-end model including, TN, channel and RN, is presented. Circuit theory is used to derive transfer functions for transmission, diffusion and reception process with ligand-receptor binding detection. These models assume that the systems are linear time-invariant. 
One of the other earliest works carried out in this context is presented in~\cite{garralda2011diffusion}. In this work, the main characteristics of DMC are explored with the use of \textit{N3Sim} simulator. Linear time-invariant property is proven to be a valid assumption for DMC scenario with a transparent receiver. 
In the later works, the same authors study the noise sources in the end-to-end MC system, including the particle sampling noise at TN, the particle counting noise related to the propagation in the channel~\cite{pierobon2011diffusion} as well as reception noise due to ligand-receptor binding at RN~\cite{pierobon2011noise}. In~\cite{pierobon2011diffusion}, the receiver is transparent, i.e. the received signal is approximated by the local concentration of molecules at the RN location. However, in \cite{pierobon2011noise} the reception process is modeled using reversible ligand-receptor kinetics but the diffusion and reception processes are analyzed independently. 
A composite MC model is introduced in~\cite{al2018modeling}. This model takes into account the issue of heterogeneity in multiple regions each with distinct diffusion properties, which can be of interest in aqueous cellular biological medium inside the human body. In~\cite{noel2016channel}, MC channel impulse response is calculated for spherical transmitters. In this paper, the simplifying assumption of point source which is very popular among MC researchers is eliminated. A channel estimation framework for diffusive MC is presented in~\cite{jamali2016channel}. The benefit of this approach is that it is not limited to a particular channel or a specific receiver type and does not require knowledge of channel parameters. However, channel estimation techniques often suffer from computational complexity issues. 

A comprehensive reactive receiver model for DMC is presented in~\cite{ahmadzadeh2016comprehensive}. In this paper, the reception mechanism at the receiver is modeled as a second-order reversible reaction. 
In a reversible second order reaction, a ligand molecule $L$ is assumed to reversibly bind to a receptor $R$ to form a ligand-receptor complex $C$ via the following reaction:
\begin{center}
$L+R \xrightleftharpoons[k_b]{k_f} C$
\end{center}
where $k_f$ and $k_b$ are the forward and backward reaction constants with units \si{m^3molecule^{-1}s^{-1}} and \si{s^{-1}}, respectively.
The impact of degradation of molecules in the channel, as well as association and dissociation processes at the receiver, are taken into account in this paper. However, the impact of receptor occupancy is neglected in this work. Modeling of the ligand-receptor interaction has also been addressed in~\cite{al2018model}. In this paper, a generalized model for the ligand-receptor protein interaction is proposed in 3-D spherically bounded, diffusive microenvironment using molecular communication paradigm. 
The impact of absorbing receivers on the number of received molecules has been investigated in~\cite{assaf2016characterizing,assaf2017influence,arifler2017monte,lu2016effect}. In~\cite{assaf2016characterizing,assaf2017influence} this is carried out with the help of \textit{N3Sim} simulator and this effect is considered as an efficient method to reduce inter-symbol interference (ISI) and increase throughput. In~\cite{arifler2017monte}, authors employ a refined Monte Carlo method to accurately simulate absorption at multiple receivers. The absorption probability is then calculated for each receiver through simulations. The authors have cross validated their simulation results for two perfectly receivers with the approximate expression used in~\cite{lu2016effect}. Although these papers study multiple receiver scenarios, a rather simplistic model is employed to characterize each receiver. The receivers are perfectly absorbing and the ligand-receptor interactions on the surface of each receiver and the receptor occupancy are neglected.
 
If we can assume that the number of ligand molecules is much higher that the number of receptors, the fluctuation in concentration of ligand molecules due to binding/releasing becomes negligible and the reversible second order reaction can be reduced to the following first order reaction:
\begin{center}
$R \xrightleftharpoons[k_b]{k_f} C$
\end{center}
In drug delivery applications, in which the absorbed molecules do not release after binding to the receptor, the following reaction will happen: 
\begin{center}
$L+R \xrightharpoonup{k_f} C$
\end{center}
\begin{center}
$R \xleftharpoondown[k_b]{} C$
\end{center}
In this paper, we aim to obtain the diffusion channel impulse response in presence of a probabilistic absorber, which is not obliged to be located at the observation point, as shown in Fig.~\ref{markovchain}. We then model a single receptor using an M/M/1/1 queue~\cite{kleinrock1976queueing}. Modeling a receptor with an M/M/1/1 queue also helps us to analyze congestion at the receiver as also suggested in~\cite{femminella2015molecular}. The congestion happens at the receiver side due to several reasons: The number of receptors at the receiver surface is limited. Therefore, if the receptor is busy at the time of ligand collision, it would be discarded. Furthermore, the ligand-receptor binding process does not occur very quickly and takes a significant time called trafficking time~\cite{femminella2015molecular}. This includes time needed for coupling with other cell surface molecules, internalization, recycling, degradation, and synthesis~\cite{atakan2016molecular}. In other words, trafficking time can be on the order of tens of seconds and can affect the reception process at the receiver. 
In M/M/1/1 queuing model, $q$ denotes the complementary blocking probability and the mean service time is the mean trafficking time. This probability depends on both the instantaneous arrival rate of molecules and the instantaneous service rate and is not fixed over time. However, after the transition time, the average $q$ reaches the steady state value, due to the constant emission rate. This steady state value of $q$ is now dependent on average arrival and service rates. This approach is beneficial in order to investigate the impact of absorption on the channel impulse response, as well as modeling the reception mechanism through ligand-receptor interaction at the receiver side using queuing theory. 

To this end, we model the system using random walk and Markov chains and make an attempt to estimate the channel impulse response by modifying the well-known channel impulse response obtained with no absorption assumption. The newly introduced coefficients $\alpha$, $\beta'$ and $\gamma$ are estimated through curve fitting. With considering the impact of absorption, we are able to obtain a more accurate estimate of the concentration at the desired location which is definitely lower than the concentration obtained by neglecting the impact of absorption. Afterward, the steady state values of absorption probability, receptor's arrival rate, and absorption rate are calculated, for a constant continuous emission of molecules. This accurate measurement can have a crucial role in drug delivery scenarios where determining the optimal rate of drug transmitting nanomachines is of great importance to prevent unwanted toxicity as well as maintaining the effectiveness of the drug delivery system. It can also help to obtain a more accurate measurement of the least effective concentration (LEC) which indicated the minimum concentration below which the drug does not have enough therapeutic effect~\cite{rathbone2013advances}.

The organization of this paper is as follows: Section II presents a random walk based channel modeling. Channel model and fitting are presented in section III. In section IV, we derive the concentration due to instantaneous and continuous emission in presence of a probabilistic absorber. Absorption probability, arrival rate and absorption rate are presented in section V. Section VI presents results and analysis. Section VII summarizes and concludes this paper with suggestion for future challenges. 
\section{Random Walk Based Channel Modeling}
Suppose we have an asymmetric 1-D random walk beginning at the origin without any absorbing barrier. Let $n_r$ represent the number of steps to the right and $n_l$ the number of steps to the left and n the total number of steps. Then:

\begin{equation}
n_r-n_l=x , n_r+n_l=n
\end{equation}

\begin{equation}
n_r=\frac{n+x}{2}, n_l=\frac{n-x}{2}
\end{equation}
Then, the probability of being at point $x$ after $n$ steps is equal to $P(x,n)$ which has a binomial distribution:
\begin{equation}
P(x,n)=\binom{n} {\frac{n+x}{2}} p^{\frac{n+x}{2}}(1-p)^{\frac{n-x}{2}}
\label{P(x,n)}
\end{equation}
where and $p$ is the probability of going one step to the right and $1-p$ the probability of going one step to the left. Note this is only possible when $n+x$ is even. This means if $x$ is even we can get there only with even number of steps and if odd with odd number of steps. 

Now suppose we have an absorber at point $m$, which absorbs the random walker with probability $q$. Then the probability $P(x,n,m)$ which shows the probability of being at $x$ after $n$ steps when we have an absorber at $m$ can be modeled as a Markov chain as shown in Fig.~\ref{markovchain}. Please note that the absorber location does not generally need to be at the intended receiver. The state probability vector at time $n$ is: 
\begin{equation}
\pi^{(n)}=\pi^{(0)}P^n , n=1,2,...
\label{state prob vector at time n}
\end{equation}
\begin{equation}
\pi^{(n)}=[\pi_j^{(n)}]
\hspace{5mm} 
\end{equation}
where $\pi_j^{(n)}$ is the probability of finding the system in state $X_n=j$ at the $n$th step and $P$ is the transition probability matrix:

\begin{equation}
\pi_j^{(n)}=P[X_n=j]
\end{equation}

\begin{equation}
P=[p_{ij}]
\hspace{5mm} i,j\in \mathbb{Z}\\ 
\end{equation}
 A 1-D random walk with a partly reflecting partly absorbing barrier is also investigated in~\cite{lauwerier1952linear}. However, it fails to obtain a simple expression for the distribution of the particle location. Moreover, a 1-D random walk in presence of a totally reflecting barrier is analyzed in~\cite{orlowski20031}. In~\cite{andrews2009accurate}, particle-based simulations are presented for modeling interaction between molecules and planar surfaces, namely adsorption, desorption and partial transmission. The concentration derived in this paper is the time-varying absorbed concentration and does not include the probability function of particle location in presence of an individual receptor.
\subsubsection{Probabilistic absorber located between origin and destination}

In the case when absorber is located between starting point and ending point, the probability of $P(x,n,m)$ 
for $x>0$ can be expressed as the following:
\begin{equation}
P(x,n,m)=p^{x+\frac{\Delta}{2}}(1-p)^{\frac{\Delta}{2}}\sum_{i=0}^{\frac{\Delta}{2}} b_i(1-q)^{\frac{\Delta}{2}+1-i}
\label{Analyticmlx}
\end{equation}
where $\Delta$ is $n-x$. This probability is independent of the absorber location. Because, the number of loops at the position of absorber, which represents the number of crossings, does not change with moving the absorber location between the origin and destination. In other words, there is a symmetry among paths.

Table~\ref{bi_coeff}, represents $b_i$ coefficients for different values of $\frac{\Delta}{2}$. $b_i$ coefficients indicate the number of paths with $\frac{\Delta}{2}+1-i$ crossings through absorber location, without being absorbed. This number lies between a minimum crossing of one and a maximum crossing of $\frac{\Delta}{2}+1$. Therefore, $b_0$ indicates the number of paths with maximum crossing through the absorber location.

In the case of $\frac{\Delta}{2}=0$, in which the number of steps equals the distance between the origin and destination, there is one possible combination to reach $x$ (and with a maximum crossing of one) and thus $b_0=1$. If $\frac{\Delta}{2}=1$ the random walker can cross each point either once or twice. The mere two possible combinations of twice crossing at each point including absorber are depicted in Fig. \ref{n-x=2}. The total number of combinations is $\binom{n}{\frac{n+x}{2}}$ which becomes $\binom{\Delta+x}{\frac{\Delta+2x}{2}}$ in terms of $\Delta$. Therefore the total number of combinations will be $x+2$ in this case and therefore the total number of combinations with one crossing is $x+2-2=x$. Now consider $\frac{\Delta}{2}=2$. In this case, the random walker can cross each point one to three times. The four possible combinations to reach $x$ with three crossings is depicted in Fig. \ref{n-x=4}. Therefore, $b_i$ coefficients are defined as follows:

\begin{figure}
\centering
  \begin{subfigure}{0.5\textwidth}
    \includegraphics[width=\textwidth]{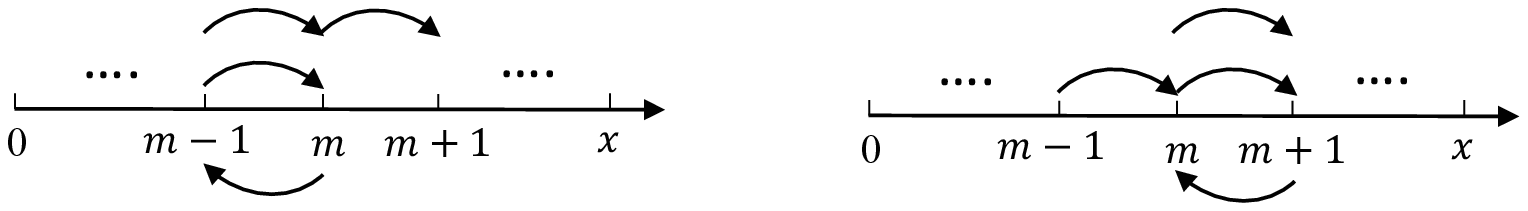}
    \caption{}
    \label{n-x=2}
  \end{subfigure}
  
  \begin{subfigure}{0.4\textwidth}
    \includegraphics[width=\textwidth]{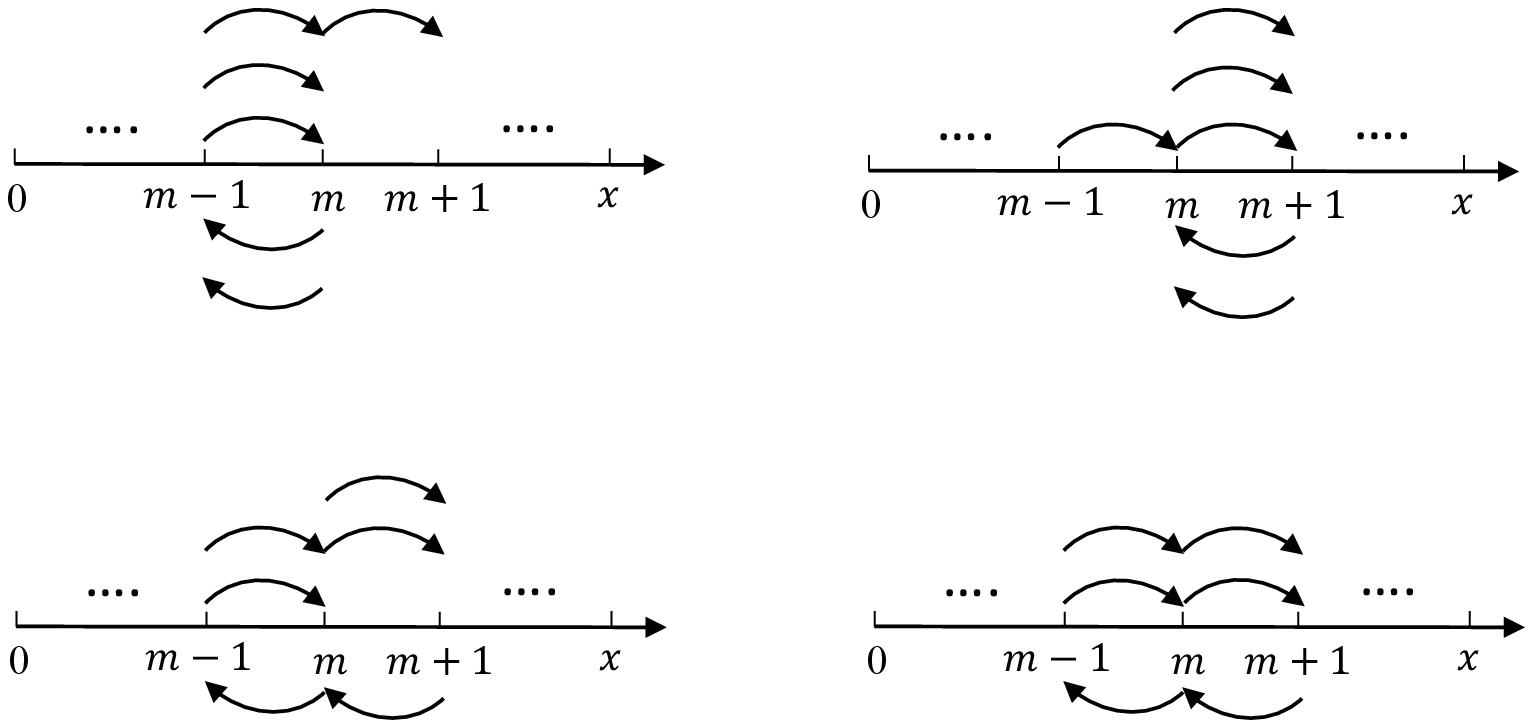}
    \caption{}
    \label{n-x=4}
  \end{subfigure}
    
  \caption{Different possible combinations of crossing the absorber location for (a) $\frac{\Delta}{2}=1$, (b) $\frac{\Delta}{2}=2$}
  \centering
  \label{crossing}
\end{figure}

\begin{table*}
\centering
\caption{Representation of $b_i$ coefficients for $0<m<x$} 
\begin{tabular}{l*{5}{c}}
              & $b_0$ & $b_1$ & $b_2$ & $b_3$& $b_4$\\
\hline
$\frac{\Delta}{2}=0$ & 1 &  & \\
$\frac{\Delta}{2}=1$            & 2 & $x$ & & &\\
$\frac{\Delta}{2}=2$           & 4 & $2(x+1)$ & $\frac{1}{2!}x(x+3)$ & &\\
$\frac{\Delta}{2}=3$     & 8 & $4(x+2)$ & $\frac{2}{2!}(x+1)(x+4)$ & $\frac{1}{3!}x(x+4)(x+5)$ & \\
$\frac{\Delta}{2}=4$ & 16 & $8(x+3)$ & $\frac{4}{2!}(x+2)(x+5)$ & $\frac{2}{3!}(x+1)(x+5)(x+6)$ & $\frac{1}{4!}x(x+5)(x+6)(x+7)$ \\
\end{tabular}
\label{bi_coeff}
\end{table*}



\begin{equation}
b_i=2^{\frac{\Delta}{2}-i}\binom{x+\frac{\Delta}{2}+i}{x+\frac{\Delta}{2}}\frac{x+\frac{\Delta}{2}-i}{x+\frac{\Delta}{2}+i}
\label{b_i}
\end{equation}





\subsubsection{Probabilistic absorber located on the origin or destination}

Suppose the absorber is located on origin or destination and the absorption does not happen when the particle starts at the origin and when it reaches the destination at the last step. Then the probability is similar to the previous section in which $0<m<x$ except for one less absorption. Thus, $P(x,n,m)$ for $x>0$ is as follows:
\begin{equation}
P(x,n,m)=\frac{1}{1-q}p^{\frac{\Delta}{2}+x}(1-p)^{\frac{\Delta}{2}}\sum_{i=0}^{\frac{\Delta}{2}} b_i(1-q)^{\frac{\Delta}{2}+1-i}
\label{Analyticmex}
\end{equation}
\begin{gather*}
=p^{\frac{\Delta}{2}+x}(1-p)^{\frac{\Delta}{2}}\sum_{i=0}^{\frac{\Delta}{2}} b_i(1-q)^{\frac{\Delta}{2}-i}
\end{gather*}
in which $b_i$ can be found through Eq.~\eqref{b_i}. 

\subsubsection{Probabilistic absorber located outside the range of origin or destination}

In this case, the impact of absorber's location on the observation point has symmetry with respect to point $x/2$. Thus for $m>x>0$ we have:
 
\begin{equation}
P(x,n,m)=P(x,n,x-m)
\end{equation}

\begin{equation}
\begin{split}
P(x,n,m)&=\\
&\left\{
	\begin{array}{ll}
	P(x,n)	  &  n<|2m-x|\\	p^{\frac{\Delta}{2}+x}(1-p)^{\frac{\Delta}{2}}[h+\\
\sum_{i=0}^{\frac{\Delta}{2}-\eta} c_i(1-q)^{\frac{\Delta}{2}-\eta+1-i}] & n\geq |2m-x|
	\end{array}
\right.
\end{split}
\label{Analyticmgx}
\end{equation}
Eq.~\eqref{Analyticmgx} indicates that the particle does not reach the absorber location, if the number of steps is less than $|2m-x|$. Coefficients $c_i$ are defined as follows:


\begin{equation}
c_i=2^{\frac{\Delta}{2}-\eta-i}\binom{x+\frac{\Delta}{2}+\eta+i}{x+\frac{\Delta}{2}+\eta}\frac{x+\frac{\Delta}{2}+\eta-i}{x+\frac{\Delta}{2}+\eta+i}
\end{equation}
and $\eta$ shows distance from destination (origin):



\begin{equation}
\eta=
\left\{
	\begin{array}{ll}
	m-x	  & \mbox{if } m>x\\
	-m & \mbox{if } m<0 
	\end{array}
\right.
\end{equation}
$h$ represents the coefficient for zero number of absorber crossing 
and can be written as follows:

\begin{equation}
h=\binom{n}{x+\frac{\Delta}{2}} -\sum_{i=0}^{\frac{\Delta}{2}-\eta}c_i 
\end{equation}
$h$ can be represented in the following explicit form:

\begin{equation}
h=
\left\{
	\begin{array}{ll}
	\binom{n}{\frac{\Delta}{2}}-\binom{n}{\frac{\Delta}{2}-\eta}  & \mbox{if } \eta \leq \Delta/2\\
	\binom{n}{\frac{\Delta}{2}} & \mbox{if }  \eta > \Delta/2 
	\end{array}
\right.
\end{equation}

For higher dimensions, $P(X,n,m)$, which represents the probability of finding the particle at the Cartesian coordinate of $X$ after $n$ steps, in presence of a probabilistic absorber at the Cartesian coordinate of $m$ can be obtained through similar Markov chain aproach. However, it is difficult to derive analytic expressions similar to Eqs.~\eqref{Analyticmlx}, \eqref{Analyticmex} and \eqref{Analyticmgx} in these cases. In the next secation, we suggest a fitting probability function for these Markov-based results.  
\section{Channel Model and Fitting} 
In this section, we use the probability function of particle location in an environment with no absorber, represented in~\eqref{pdf}, to develop a diffusion-based molecular channel model which consists of a probabilistic absorber. 
\begin{equation}
P(X,t)=\frac{2}{(4\pi D t)^{d/2}}e^{\frac{-{r}^2}{4Dt}}
\label{pdf}
\end{equation}
in which $D=\frac{\delta^2}{2d\tau}$ is the diffusion coefficient, $d$ is the dimension of the diffusion environment, $\delta$ is the step length, $\tau$ is the time step, $X$ is the Cartesian coordination of the observation point and $r$ is the distance of observation point from origin. Please note that this is twice the actual probabiliy to match the discrete random walk analysis. This coefficient appears in converting discrete variables to continuous ones by considering $x=2n_r-n$ and hence $dx=2dn_r$.  

In order to model the diffusion channel impulse response in presence of a probabilistic absorber with the above channel model, we introduced several adjustment parameters, namely $\alpha$, $\beta$ and $\gamma$.
Hence, $P(X,t,m)$, which represents the probability function of particle location in presence of an absorber located at $m$, can be represented as follows:
\begin{equation}
P(X,t,m)=\frac{\alpha}{(4\pi D t)^{d/2}}e^{\frac{-{r}^2}{4D\beta t}}\left(\frac{{r}^2}{Dt}\right)^{\gamma}
\label{closedform1D}
\end{equation}
in which $\alpha$, $\beta$ and $\gamma$ are the model fitting parameters and are defined as amplitude parameter, scale parameter and decrement parameter respectively. These model-fitting parameters are introduced to compensate the difference between channel impulse response with and without an absorbing point. The extra term, $(\frac{{r}^2}{Dt})^{\gamma}$, is added to capture the impact of absorber on the decay of the tail of the response. The channel fitting approach has also been used in~\cite{farsad2014channel} in two extreme cases of our problem, i.e., $q=0$ and $q=1$ to match corrected forms of diffusion equation and first hitting probability to experimental data, respectively. 

To find the model parameters, we use the nonlinear least squares curve-fitting technique. Assuming $K$ observation, the parameter estimation problem with three parameters is formulated as follows:
\begin{equation}
\begin{aligned}
&\underset{\alpha,\beta,\gamma}{\text{arg~min}}
& &\sum_{i=1}^{K} (P(X,t_i,m)-P(X,n_i,m))^2
\label{optimization problem}
\end{aligned}
\end{equation}
In the next section, we use Eq.~\eqref{closedform1D} to obtain the concentration due to the instantaneous and continuous emission in presence of a probabiistic absorber.






\section{Concentration due to instantaneous and continuous emission in presence of  probabilistic absorber}
According to Eq.~\eqref{closedform1D}, if we assume that $N$ molecules are released instantaneously at the origin of a coordinate system, the molecule concentration $C(X,t,m)$ is expressed as follows:
\begin{equation}
C(X,t,m)=\frac{\alpha N}{(4\pi D t)^{d/2}}e^{\frac{-{r}^2}{4D\beta t}}\left(\frac{{r}^2}{Dt}\right)^{\gamma}
\label{concentration1D}
\end{equation}

If we assume that molecules are released at a rate $Q(t)$, instead of instantaneous emission, then the concentration at distance $x$ and time $t$ is obtained by integrating~\eqref{concentration1D} as follows\cite{atakan2016molecular,bossert1963analysis}:

\begin{equation}
\begin{split}
C(X,t,m)&=\\
\int_{0}^{t} &\frac{\alpha Q(t_0)}{(4\pi D (t-t_0))^{d/2}}e^{\frac{-{r}^2}{4D\beta (t-t_0)}}\left(\frac{{r}^2}{D(t-t_0)}\right)^{\gamma}dt_0
\end{split}
\end{equation}
The continuous emission of molecules is of interest in healthcare applications of molecular communication such as drug delivery in which an effective concentration of medication is required to be present at the target site over the period of treatment.
With substitution $v=\frac{r^2}{4D\beta (t-t0)}$ and assuming constant emission rate, $Q(t)=Q$, the above integral is simplified to:
\begin{equation}
\begin{split}
C(X,t,m)&=\\
&\frac{\alpha Q}{(4\pi)^{d/2} D}r^{2-d}(4 \beta)^{\gamma-\frac{2-d}{2}} \Gamma\left(\gamma-\frac{2-d}{2},\frac{r^2}{4D\beta t}\right)
\label{Continuousemission1D}
\end{split}
\end{equation}
in which $\Gamma(\cdot,\cdot)$ is the incomplete gamma function and is defined as:
\begin{equation}
\Gamma(s,x)=\int_x^{\infty} t^{s-1}e^{-t}dt.
\end{equation}
Since $\Gamma(s,0)=\Gamma(s)$, the steady state concentration due to continuous emission is:
\begin{equation}
\begin{split}
C(X,m)&=\lim_{t\to\infty} C(X,t,m)\\
&=\frac{\alpha Q}{(4\pi)^{d/2} D}r^{2-d}(4 \beta)^{\gamma-\frac{2-d}{2}}\Gamma\left(\gamma-\frac{2-d}{2}\right),
\label{continuousemission1Dlim}
\end{split}
\end{equation} 
\section{Absorption probability, arrival rate and absorption rate}
\begin{figure}[htbp]
\centering
\includegraphics[width=0.4\textwidth]{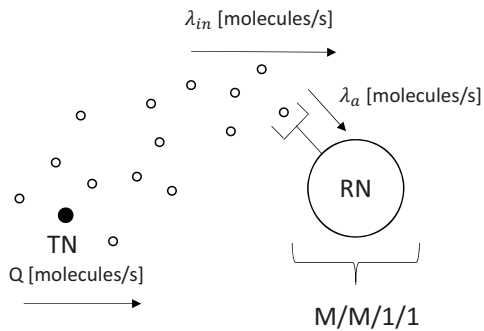}
\caption{Illustration of modeling a single receptor receiver with an M/M/1/1 queue}
\label{queue}
\centering
\end{figure}
In this section, we aim to model a receiver including one receptor, as shown in Fig.~\ref{queue}, with M/M/1/1 queue model. Assuming a continuous release rate of $Q$ at TN location, which is a point source, the absorption rate can be calculated as follows:
\begin{equation}
\lambda_a=q\lambda_{in}   
\end{equation}
in which $q$ is the steady state absorption probability and $\lambda_{in}$ is the steady state arrival rate at RN location. Starting from $\lambda_{in}$, derived from Eq.~\eqref{continuousemission1Dlim} for $x=m$ and the initial value of $q=1$, indicating a free receptor, and according to~\cite{bertsekas1992data} for an M/M/1/1 queue, we have:
\begin{equation}
q=1-p_b=\frac{\mu}{\lambda_{in}+\mu}
\label{q}
\end{equation}
in which $p_b$ stands for receptor's blocking probability. Service time has an exponential distribution with rate parameter $\mu$ in which $T_{trafficking}=1/\mu$ is the mean service time. Therefore, according to Eqs.~\eqref{continuousemission1Dlim} and~\eqref{q} the steady state values of $q$, $\lambda_{in}$ and $\lambda_a$ can be obtained.
\section{Results and Analysis} 
\begin{figure}
\centering
\includegraphics[width=0.45\textwidth]{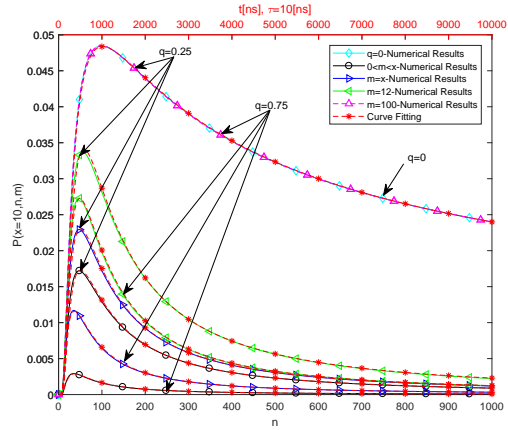}
\caption{Probability of reaching $x=10$ in presence of a probabilistic absorber for several values of $q$ in 1-D environment, where the upper axis  in red represents the continuous variable $t$ and the lower one indicates the discrete variable $n$.}
\label{Pxmn123OriginalVsClosedFormEst}
\centering
\end{figure}

\begin{figure*}
  \begin{subfigure}[b]{0.32\textwidth}
    \includegraphics[width=\textwidth]{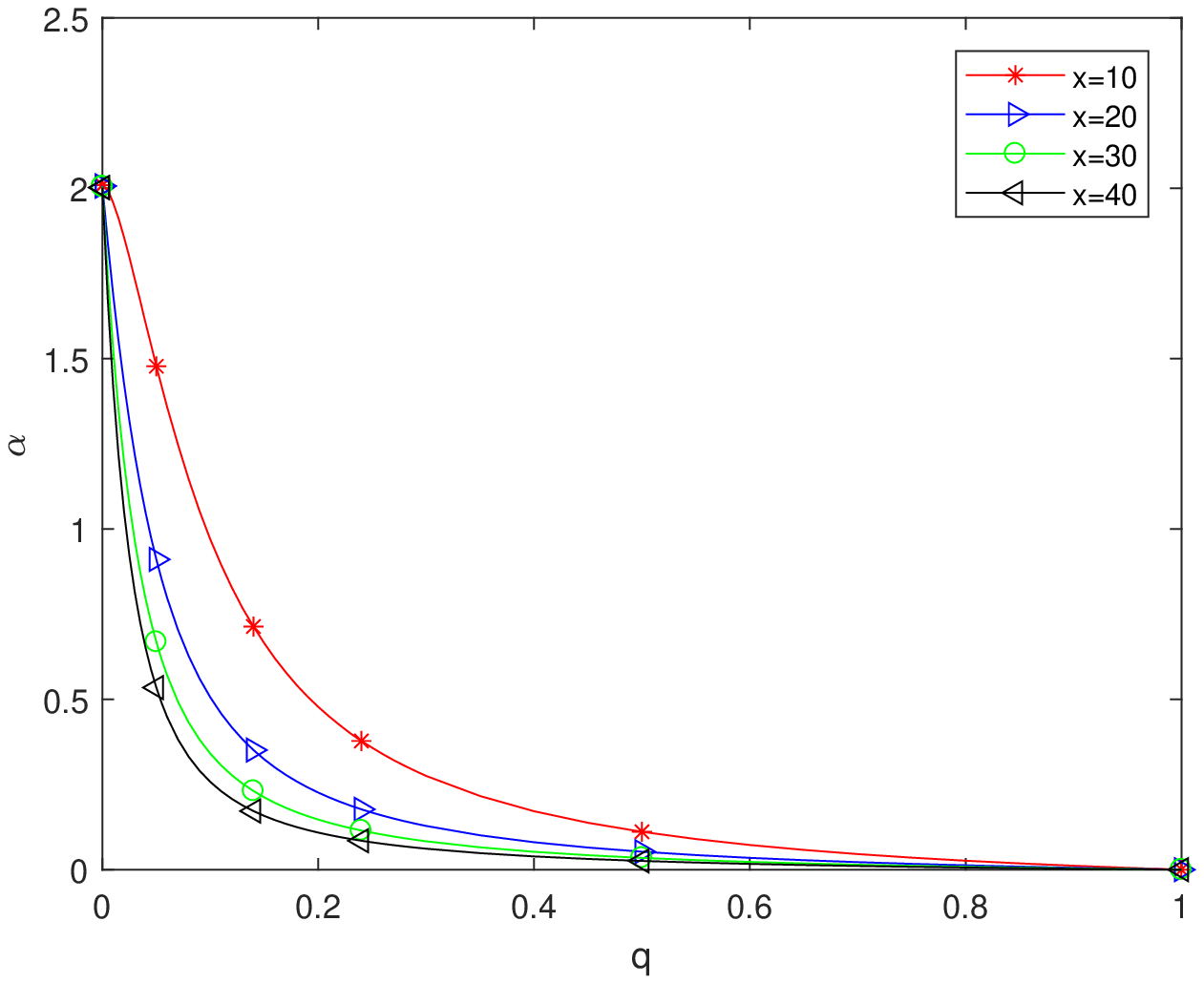}
    \caption{}
    \label{AlphaVsQ_AbsorberInside}
  \end{subfigure}
  \begin{subfigure}[b]{0.32\textwidth}
    \includegraphics[width=\textwidth]{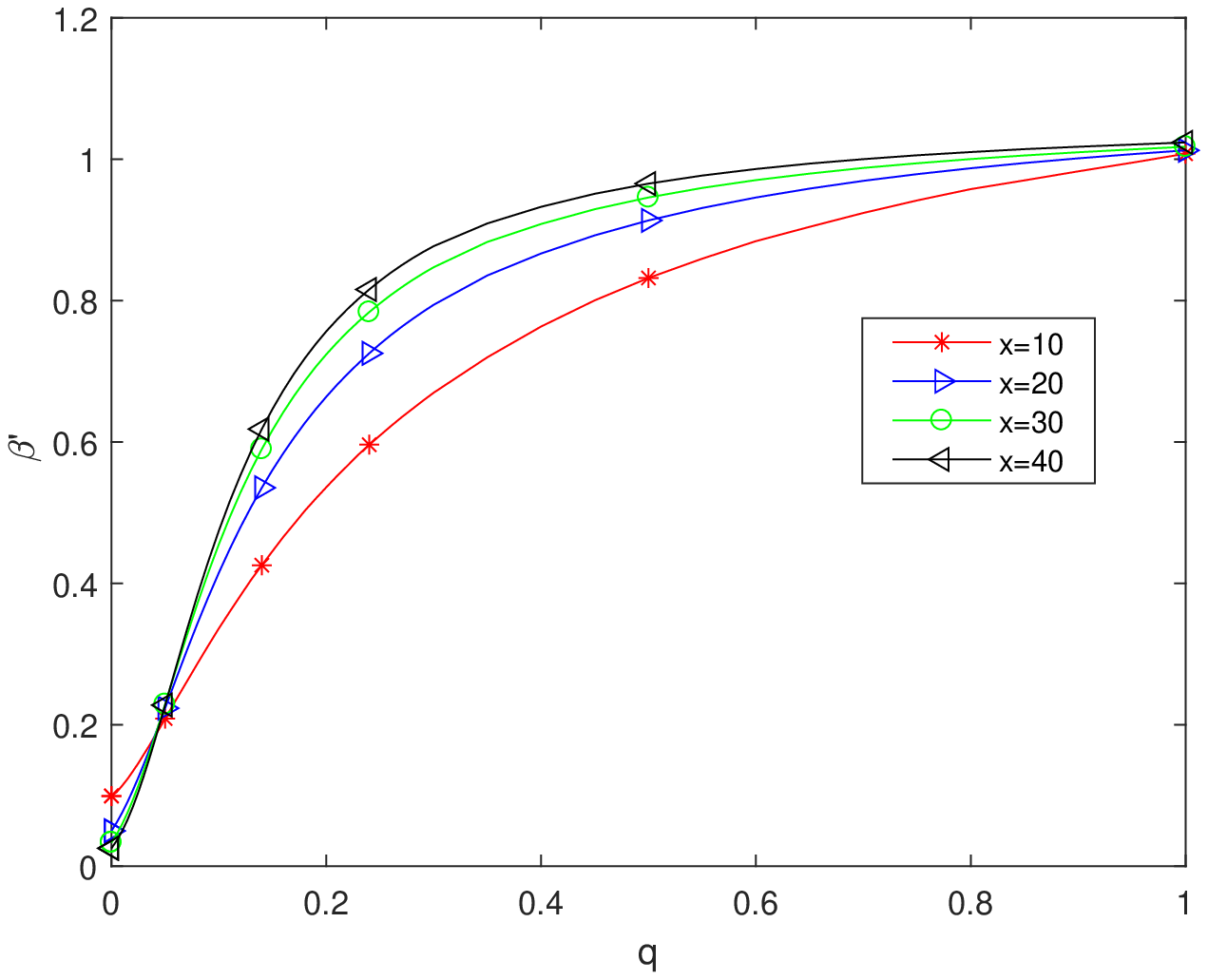}
    \caption{}
    \label{BetaVsQ_AbsorberInside}
  \end{subfigure}
  \begin{subfigure}[b]{0.32\textwidth}
    \includegraphics[width=\textwidth]{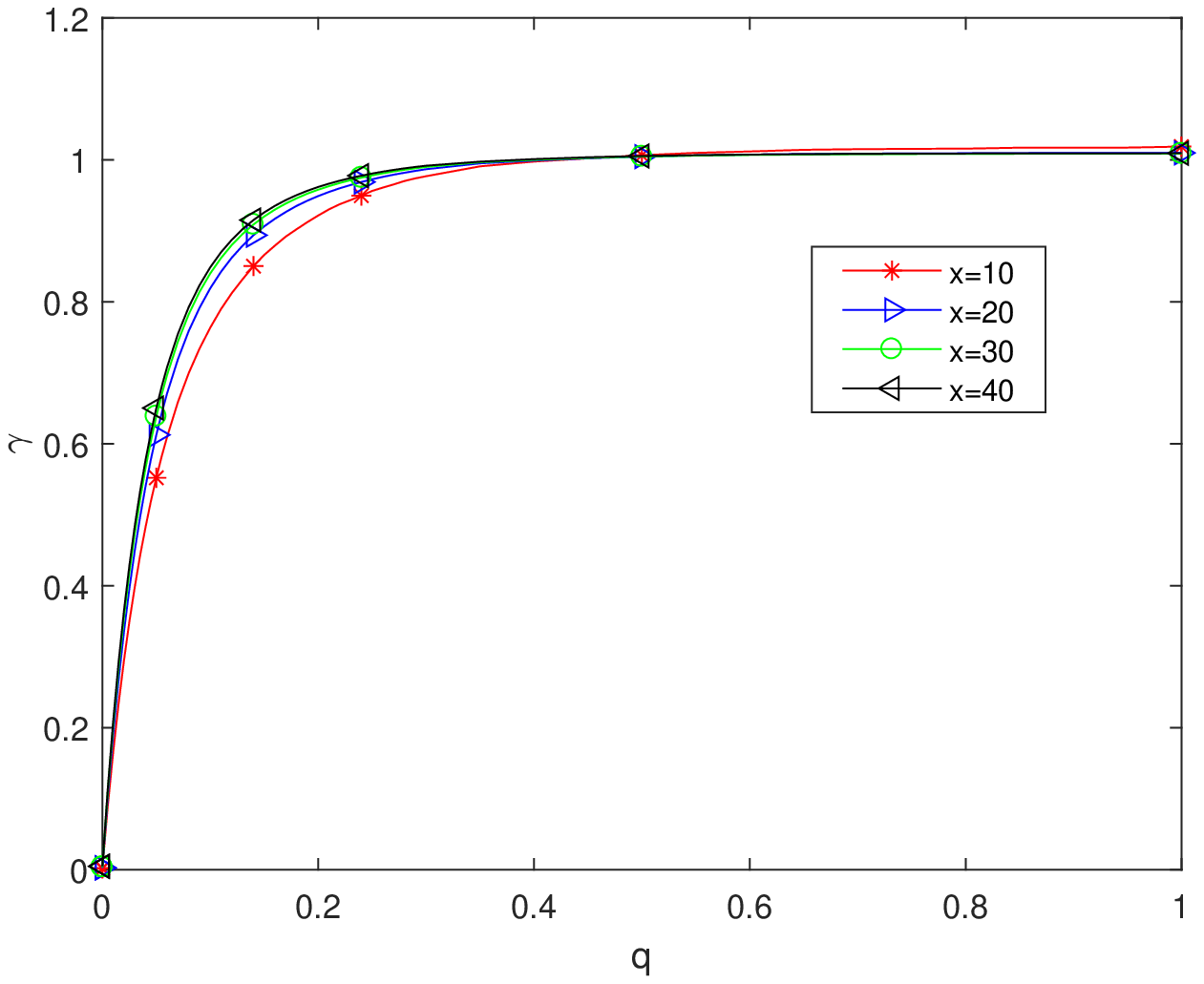}
    \caption{}
    \label{GammaVsQ_AbsorberInside}
  \end{subfigure}
  \caption{Parameter $\alpha$,~$\beta'$ and $\gamma$ as a function of $q$ for $0<m<x$ and 1-D environment}
\end{figure*}
\begin{figure*}
  \begin{subfigure}[b]{0.32\textwidth}
    \includegraphics[width=\textwidth]{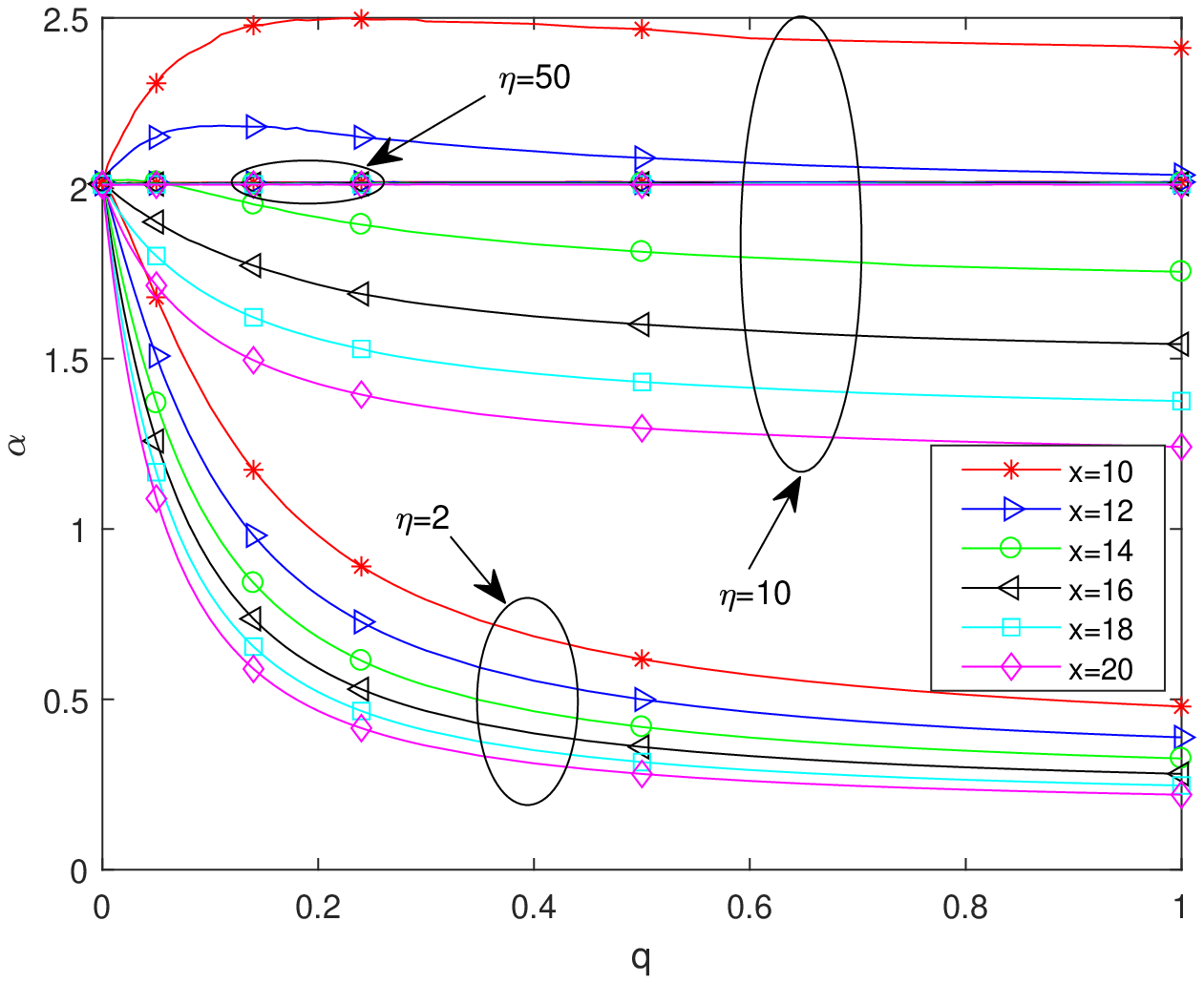}
    \caption{}
    \label{AlphaVsQ_AbsorberOutsideMu=2ConsistentMarker}
  \end{subfigure}
  \begin{subfigure}[b]{0.32\textwidth}
    \includegraphics[width=\textwidth]{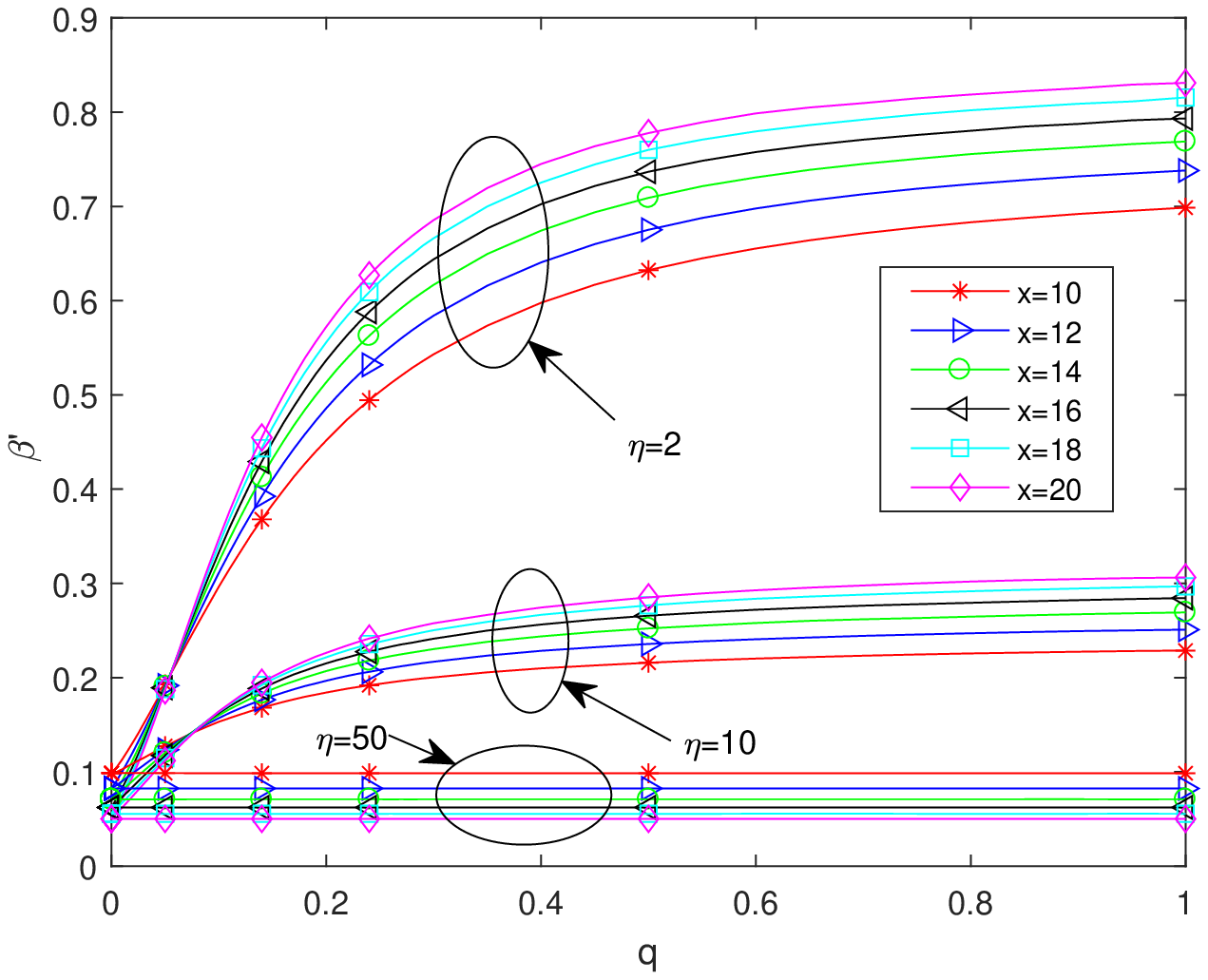}
    \caption{}
    \label{BetaVsQ_AbsorberOutsideMu=2ConsistentMarker}
  \end{subfigure}
  \begin{subfigure}[b]{0.32\textwidth}
    \includegraphics[width=\textwidth]{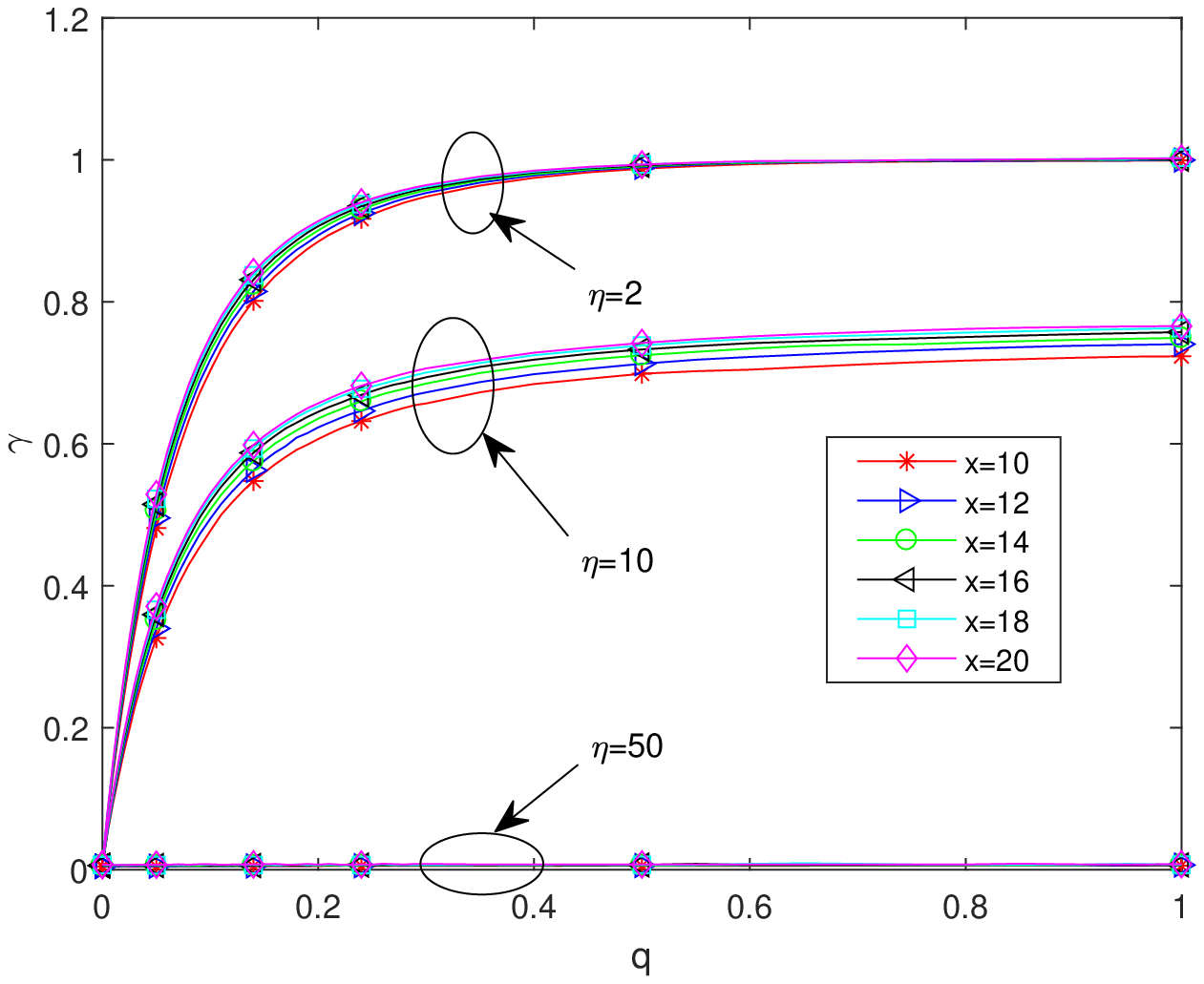}
    \caption{}
    \label{GammaVsQ_AbsorberOutsideMu=2ConsistentMarker}
  \end{subfigure}
  \caption{Parameters $\alpha$,~$\beta'$ and $\gamma$ as a function of $q$ for $m>x$ in 1-D environment}
\end{figure*}
In this section, we provide the results and analysis of our channel modeling approach.
Fig.~\ref{Pxmn123OriginalVsClosedFormEst} plots the probability functions of $P(x,n,m)$ and $P(x,t,m)$, obtained from Markov-based results and curve-fitting, respectively. In this figure, the probability functions are compared for different absorber's locations and various values of $q$ in 1-D environment, where the upper axis in red represents the continuous variable, $t$,  and the lower one indicates the discrete variable, $n$. As we see the lowest reaching probability belongs to the case where the absorber is located midway between the origin and destination. The probability gets higher if it is located on the boundary points or outside. Fig.~\ref{Pxmn123OriginalVsClosedFormEst} also shows that for a specified location of the absorber, the probability of reaching $x=10$ is reduced as the absorption probability is increased. Furthermore, we can observe that when the absorber is located outside the origin-destination interval the probability of reaching rises as the distance between the absorber and destination is increased. The probability function is also shown for a relatively large distance of $\eta=90$. In this case, the curve is almost indistinguishable from the no absorber curve. In other words, the impact of the absorber is negligible in this case. It is obvious that the absorption probability does not alter the probability function in this case. We also note the excellent match between the Markov-based and curve-fitting results.

The parameters $\alpha$, $\beta'$ and $\gamma$ for $0<m<x$ are plotted in Fig.~\ref{AlphaVsQ_AbsorberInside}-\ref{GammaVsQ_AbsorberInside} as a function of $q$, respectively. Here, we define $\beta'=\beta x^{\gamma-1}$. 
Parameters $\alpha$, $\beta'$ and $\gamma$ for $m=0$ or $m=x$ are very similar to the former case values.
Parameters $\alpha$, $\beta'$ and $\gamma$ for $m>x$ are shown in Fig.~\ref{AlphaVsQ_AbsorberOutsideMu=2ConsistentMarker}-\ref{GammaVsQ_AbsorberOutsideMu=2ConsistentMarker} for three different values of $\eta$ and some different destination locations.
It can be seen from the figures that parameters generally depend on the absorption probability $q$, the absorber location $m$ and destination location $x$. Also, we can verfy that for $q=0$, Eq.~\eqref{closedform1D} will reduce to the original $P(x,t)$ of Eq.~\ref{pdf}, which corresponds to $\alpha=2$, $\beta=1$ and $\gamma=0$.
It is also worth to note that the 1-D probability density function of the particle location for a specified location and a fully absorbing barrier located there, i.e. $q=1$ and $m=x$, corresponds to the well-known first-passage probability of Eq.~\eqref{fhit}. 
\begin{equation}
f(t)=\frac{2x}{\sqrt{4\pi Dt^3}}e^{\frac{-x^2}{4Dt}}
\label{fhit}
\end{equation}

\begin{figure}
\centering
\includegraphics[width=0.5\textwidth]{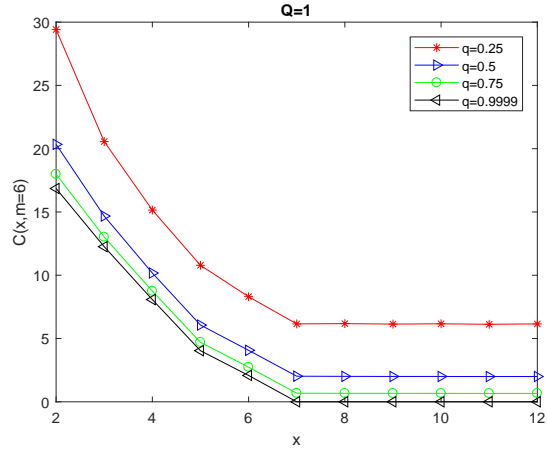}
\caption{Steady state concentration as a function of $x$ for absorber located at $m=6$ and various values of $q$, $Q=1$~molecule/ns in 1-D environment}
\label{Cx}
\centering
\end{figure}

The steady state concentration at $x$ in presence of a probabilistic absorber located at $m=6$ is shown in Fig.~\ref{Cx}. Since the concentration can be calculated for absorption probabilities which result in $\gamma>0.5$, according to Eq.~\eqref{continuousemission1Dlim}, Fig~\ref{Cx} is plotted for $q \geq 0.25$. The steady state concentration at each location is decreased as the absorption probability is increased. The concentration at locations for which the absorber locates outside the origin-destination interval depends on the distance of absorber from the destination and is declined as the distance is decreased. However, when the absorber is located midway, the concentration remains constant as the destination location, and hence the distance between the absorber and destination, vary. 
It can also be observed that when $q \to 1$, the concentration tends to zeros, as in 1-D environment the only possible way for the particle to reach destination is to cross the absorber location. 

\begin{figure}
\centering
\includegraphics[width=0.5\textwidth]{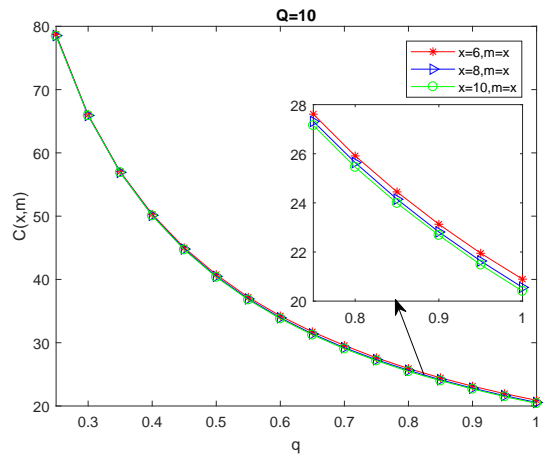}
\caption{Steady state concentration as a function of $q$ for $x=m$ in 1-D environments}
\label{Cq}
\centering
\end{figure}

Fig.~\ref{Cq} shows the steady state concentration at the location of a probabilistic absorbing receiver in terms of the absorption probability $q$ and for $Q=10$~molecule/ns. It is clear that increasing the absorption probability, leads to a significant reduction in concentration at the receiver's location. For instance, assuming a fully absorbing receiver, i.e. $q=1$, the concentration is 20\%-30\% of the case which the receiver absorbs the particle with $q=0.25$. Please note that here we have obtained the concentration present at the receiver location and is different from the absorption rate of the receiver which will be calculated later.

\begin{figure*}[htbp!]
\centering
  \begin{subfigure}{0.32\textwidth}
    \includegraphics[width=\textwidth]{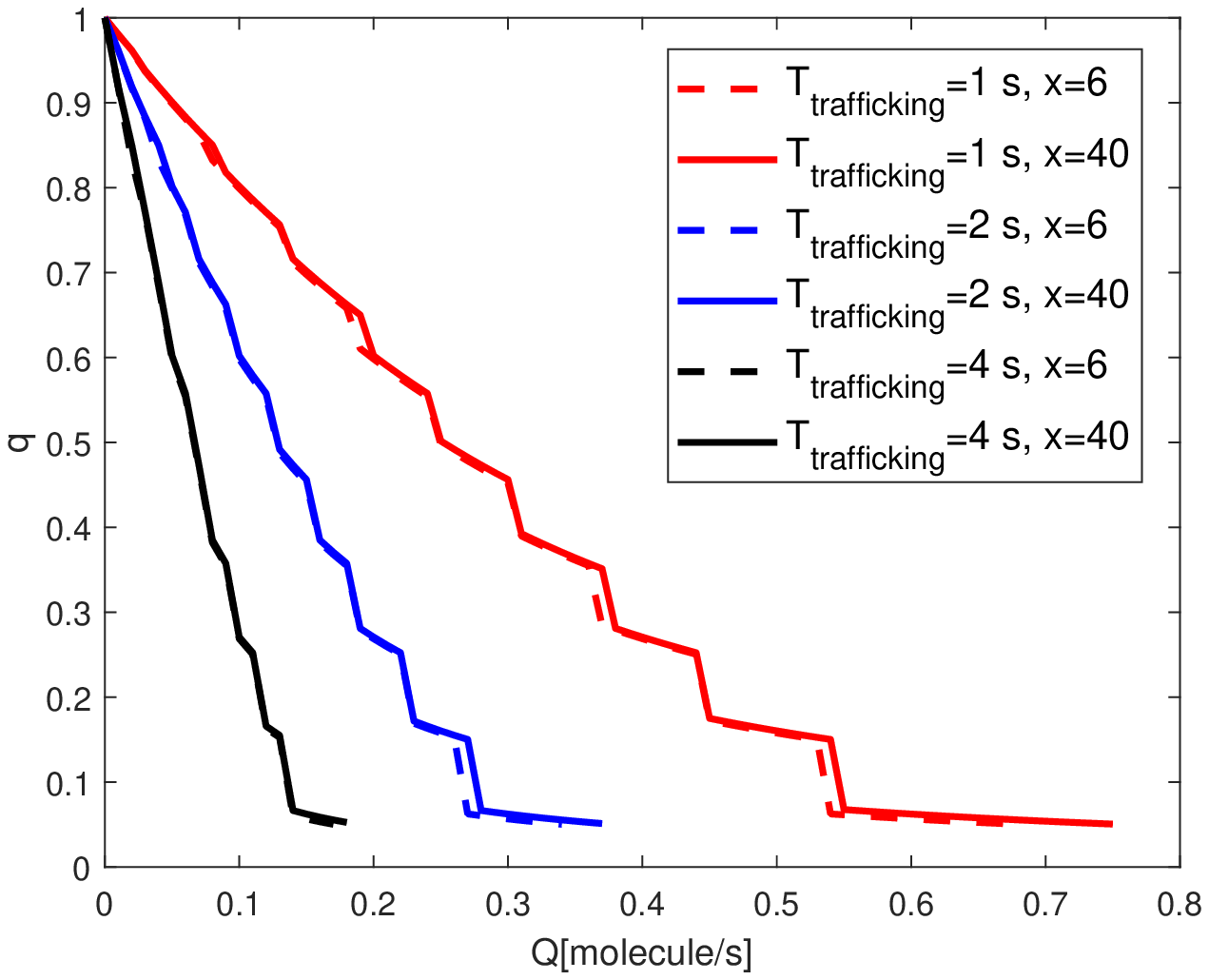}
    \caption{}
    \label{MM111Da}
  \end{subfigure}
  \begin{subfigure}{0.32\textwidth}
    \includegraphics[width=\textwidth]{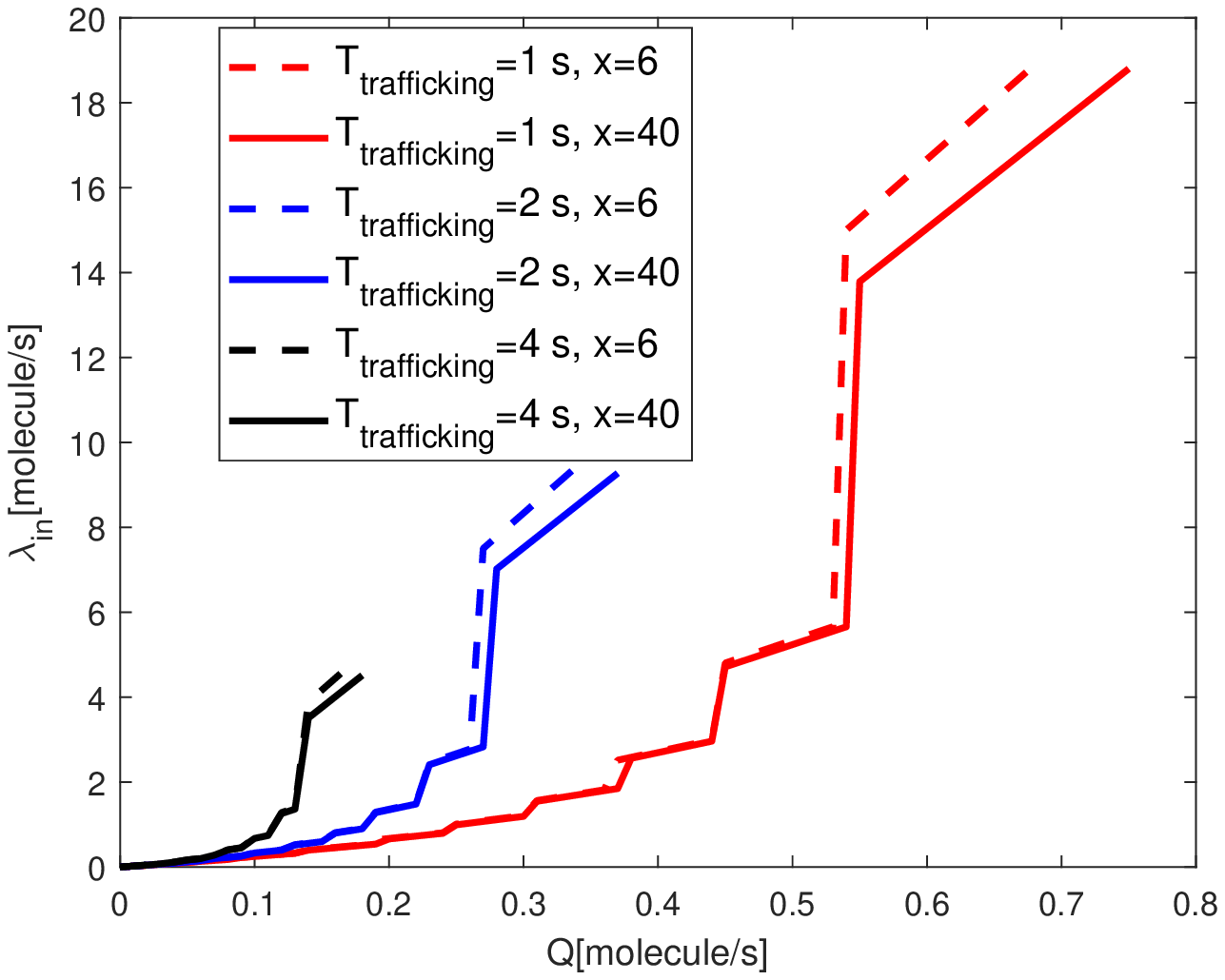}
    \caption{}
    \label{MM111Db}
  \end{subfigure}
  \begin{subfigure}{0.32\textwidth}
    \includegraphics[width=\textwidth]{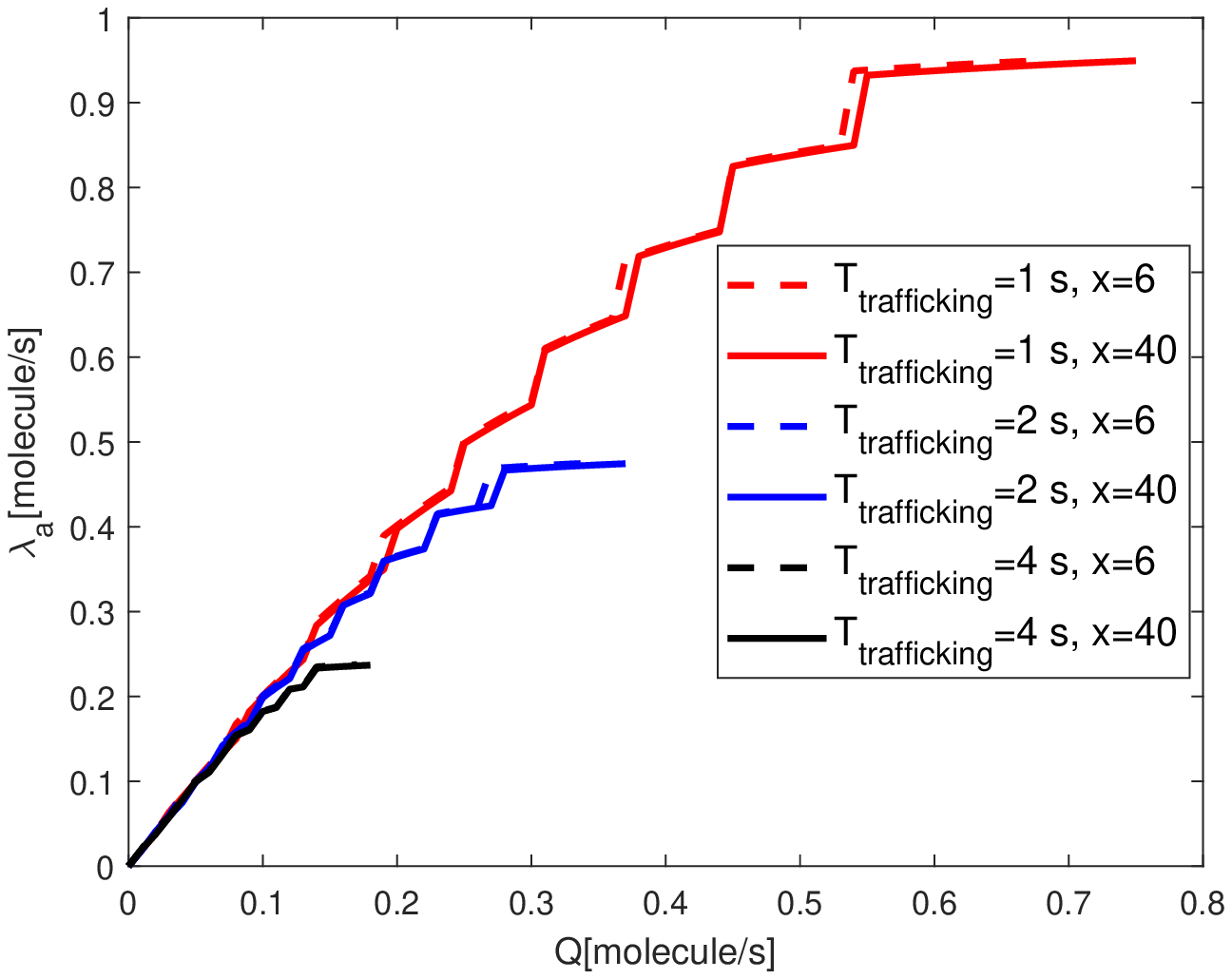}
    \caption{}
    \label{MM111Dc}
  \end{subfigure}
  \caption{Absorption probability $q$, arrival rate $\lambda_{in}$ and absorption rate $\lambda_a$ as function of releasing rate $Q$ in 1-D environment}
  \centering
  \label{MM111D}
\end{figure*}

Fig.~\ref{MM111Da}, shows the steady state absorption probability versus $Q$ for various values of $T_{trafficking}$ and receiver locations. As expected, the congestion, i.e. $q \to 0$, happens earlier, i.e. at smaller values of $Q$, for greater values of $T_{trafficking}$. As $T_{trafficking}$ decreases, the receiver is more prompt to free receptors and thus the congestion happens at higher releasing rates.

Receptor's arrival rates for different values of trafficking time is shown in Fig.~\ref{MM111Db}. For larger values of trafficking time the system is slower. Hence the concentration is higher at the receiver location.

Finally, Fig.~\ref{MM111Dc} shows the receptor's absorption rate as a function of releasing rate. The absorption rate rises as the trafficking time declines and becomes saturated at the value of $1/T_{trafficking}$. It can be  observed that for a fixed value of $T_{trafficking}$ and $Q$, as the receptor gets away from the TN location, the absorption probability grows while $\lambda_{in}$ declines and $\lambda_a$ saturates at higher levels of $Q$.  

\begin{figure}
\centering 
\includegraphics[width=0.5\textwidth]{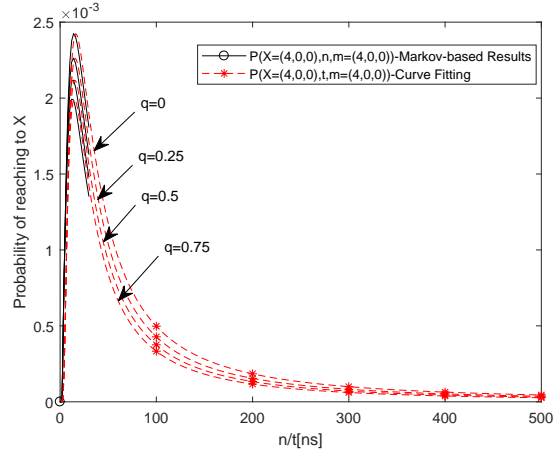} 
\caption{The probability of reaching $X=(4,0,0)$ in presence of a probabilistic absorber at $m=X$ for various values of absorption probability in 3-D environment}
\label{prob3D_diffqs} 
\centering 
\end{figure}

If the probabilistic absorber is located at the receiver site, $P(X,n,m)$ and $P(X,t,m)$ are shown in Fig.~\ref{prob3D_diffqs} in 3-D environment.  

  %
    %

\begin{figure}
\centering 
\includegraphics[width=0.5\textwidth]{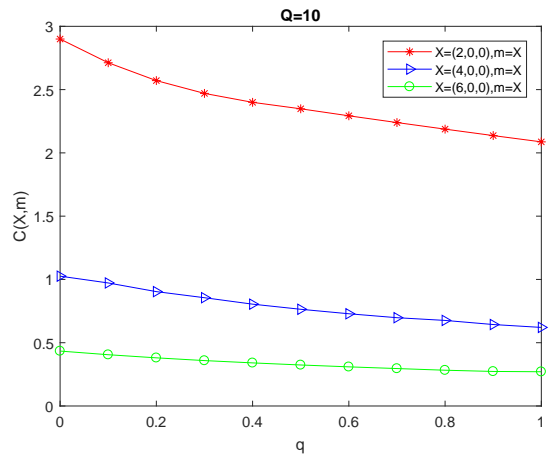}
\caption{Steady state concentration at $X=m$ as a function of $q$ in 3-D environments} 
\label{Cq3D} 
\centering 
\end{figure}

The steady state concentration at the location of a probabilistic absorbing receiver is shown in Fig.~\ref{Cq3D} as a function of $q$ for a continuous release rate of $Q=10$~molecule/ns. It can be seen that with growing the absorption probability to one, the concentration at the receiver location declines to about 60\%-70\% of the zero absorption case. It can be seen that the impact of the absorber is less significant as the environment dimensions grows. This is because the number of paths from origin to destination which do not cross the absorber is increased as the dimension grows.

\begin{figure*}
\centering
  \begin{subfigure}{0.32\textwidth}
    \includegraphics[width=\textwidth]{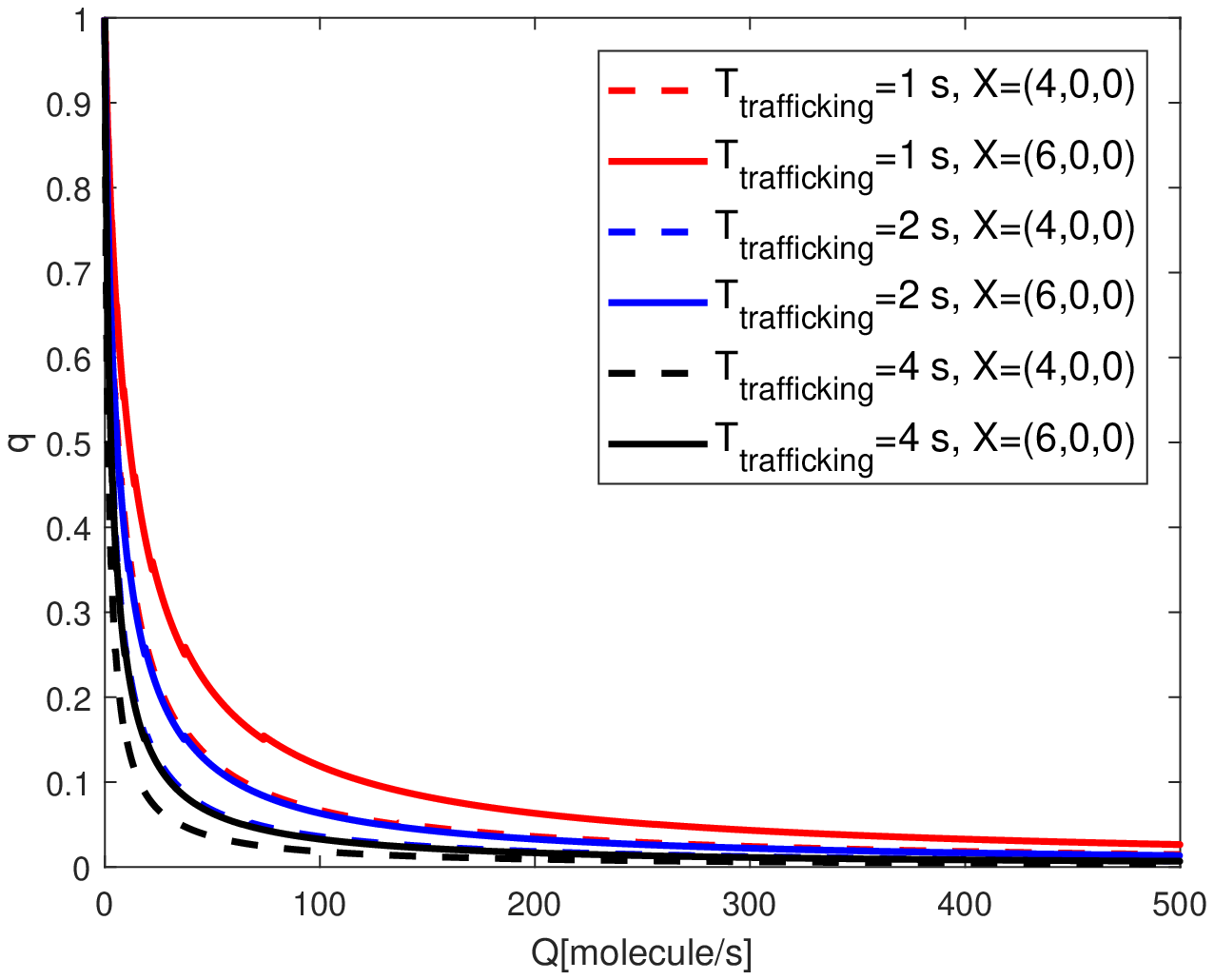}
    \caption{}
    \label{MM113Da}
  \end{subfigure}
  \begin{subfigure}{0.32\textwidth}
    \includegraphics[width=\textwidth]{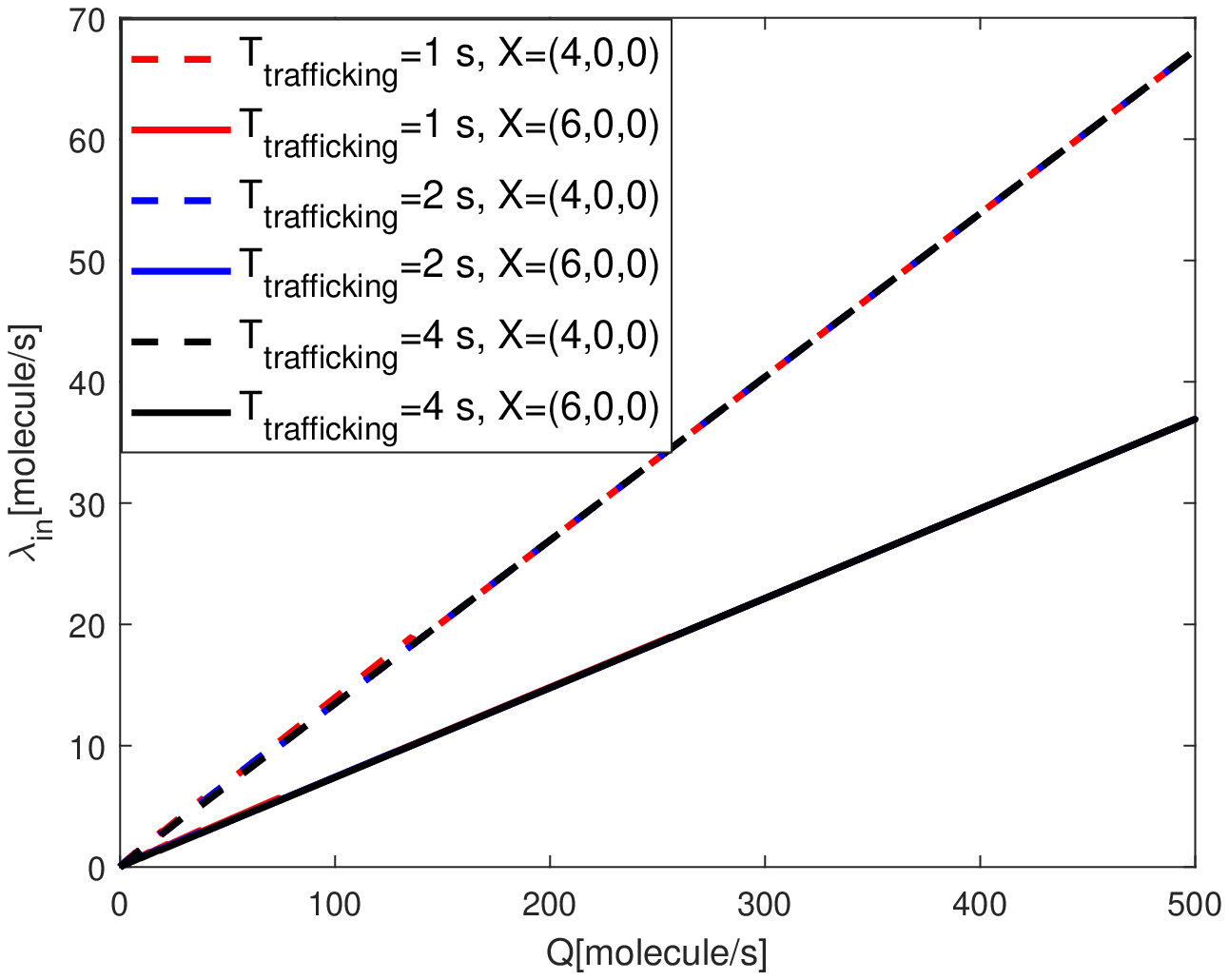}
    \caption{}
    \label{MM113Db}
  \end{subfigure}
  \begin{subfigure}{0.32\textwidth}
    \includegraphics[width=\textwidth]{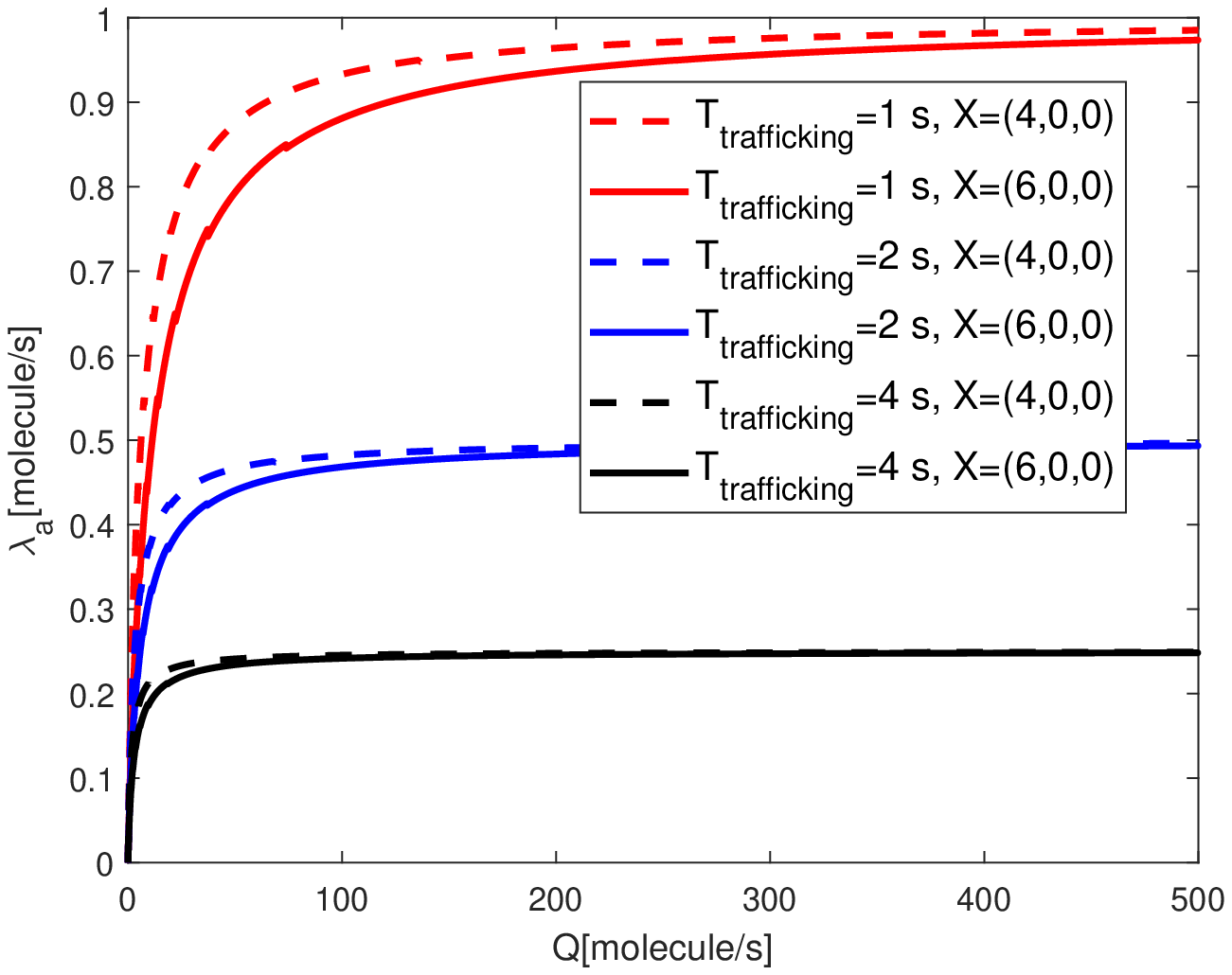}
    \caption{}
    \label{MM113Dc}
  \end{subfigure}
  \caption{Absorption probability $q$, arrival rate $\lambda_{in}$ and absorption rate $\lambda_a$ as function of releasing rate $Q$ in 3-D environment}
  \centering
  \label{MM113D}
\end{figure*}

Fig.~\ref{MM113D}, shows the steady state absorption probability, arrival and absorption rates for 3-D environments. The same interpretations as Figs.~\ref{MM111D} hold for this case. Again, note that the system feels congested at much higher rates of TN compared to 1-D. Moreover, as we see from the figure, for the same emission rate and trafficking time, farther destinations are less congested. This occurs due to the lower concentration or arrival rates at a farther location. This difference becomes less prominent as the absorption probability tends to zero at very high emission rates. Although the absorption rates are lower for farther locations for some emission rates, it saturates to the same level as the emission rate is increased.
\section{Conclusion}
In this paper, we have modeled the DMC channel impulse response in presence of a probabilistic absorber in 1-D to 3-D environments. Generally, the absorber can be located midway, on the TN or RN location and outside the TN-RN interval. Considering the impact of the probabilistic absorber, the probability density function of the particle location, as well as concentration due to instantaneous and continuous emission, can be found precisely. Furthermore, in a particular case where the probabilistic absorber is located at RN location, which can be regarded as a model for a receptor, and assuming M/M/1/1 queue model the receptor's absorption rate can be obtained. These findings have significant importance in designing a drug delivery system and determining the optimal release rate of transmitting nanomachines. \\
In our future work, we plan to extend our analysis to multiple receptor receivers to effectively model reception mechanism in biological cells.

\nocite{*}
\bibliographystyle{IEEEtran}
\bibliography{main}

\end{document}